\documentclass[%
 reprint,
 amsmath,amssymb,
 aps,
]{revtex4-2}

\usepackage{float}
\usepackage{comment}
\usepackage{tabularx}

\usepackage[utf8]{inputenc}
\usepackage[T2A,T1]{fontenc}
\usepackage[english]{babel}

\usepackage{color}
\usepackage{ulem}
\usepackage{xcolor}
\usepackage{comment}

\usepackage{xfrac}

\usepackage{gensymb}

\usepackage[caption=false]{subfig}

\usepackage{graphicx}
\usepackage{dcolumn}
\usepackage{bm}

\begin{document}

\title{Fourier modal method for Moiré lattices}

\author{Natalia S. Karmanova}
\affiliation{Skolkovo Institute of Science and Technology, Bolshoy Boulevard 30, bld. 1, Moscow 121205, Russia}
\affiliation{Moscow Institute of Physics and Technology, Institutskiy pereulok 9, Moscow Region 141701, Russia}
\author{Ilia M. Fradkin}
\email{Ilia.Fradkin@skoltech.ru}
\affiliation{Skolkovo Institute of Science and Technology, Bolshoy Boulevard 30, bld. 1, Moscow 121205, Russia}
\affiliation{Moscow Institute of Physics and Technology, Institutskiy pereulok 9, Moscow Region 141701, Russia}
\author{Sergey A. Dyakov}
\affiliation{Skolkovo Institute of Science and Technology, Nobel Street 3, Moscow 143025, Russia}
\author{Nikolay A. Gippius}
\affiliation{Skolkovo Institute of Science and Technology, Nobel Street 3, Moscow 143025, Russia}
\date{\today}

\begin{abstract}
In recent years twisted bi-layers of 2D materials became very popular in the field due to the possibility to totally change their electronic properties by simple rotation. At the same time, in the wide field of photonic crystals, this idea still remains almost untouched, and only some particular problems were considered. One of the reasons is the computational difficulty of the accurate consideration of Moiré superlattices that appear due to the superimposition of misaligned lattices. Indeed, the unit cell of the complex lattice is typically much larger than the original crystals and requires much more computational resources for the computations. Here, we propose a careful adaptation of the Fourier modal method in the form of the scattering matrices for the description of twisted 1D gratings' stacks. Our approach allows us to consider sublattices in close vicinity to each other and account for their interaction via the near-field. In the developed numerical scheme, we utilize the fact that each sublattice is only 1D-periodic and therefore simpler than the resulting 2D superlattice, as well as the fact that even a small gap between the lattices filters out high Fourier harmonics due to their evanescent origin. This accelerates the computations from 1 up to 3 and more orders of magnitude for typical structures depending on the number of harmonics. This paves the way for rigorous study of almost any photonic crystals of the proposed geometry and demonstration of specific Moiré-associated effects.
\end{abstract}

\maketitle


\section{Introduction}

The Moiré effect, known for centuries, is a result of superimposing of similar but misaligned lattices.
Probably, everyone met Moiré fringes as an occasional effect (recognizable patterns on silk clothes and curtains), as a side-effect (prominent grids on photos of LCD screens and imposition of the computational grid on the true picture on the map graphs), but the most important are their practical applications. For example, the Moiré pattern underlies the widely-known Vernier scale, which is implemented in micrometers and other measuring devices. Also, Moiré effect is successfully applied in strain analysis \cite{kishimoto1993microcreep,post1965moire,chiang1979moire}, optical alignment \cite{drinkwater2000development,king1972photolithographic,muhlberger2007moire,li2006sub}, medicine \cite{Wang1997}, biosensors \cite{wu2016dual,Sharma2020,Zhu2018}, detection of document counterfeiting \cite{zhang1997concealed,cadarso2013high,aggarwal2006concealed,Amidror2007} and in many other spheres \cite{wu2018moire,Miao2016}. Nevertheless, the original principle of the complex superlattice design by the superimposition of identical or similar sublattices is still widely applied to observe the associated phenomena. In the last few years, there has been a breakthrough in the field of twisting 2D materials. It has been demonstrated the possibility to observe superconductivity \cite{Balents2020,Chen2019,Cao2018}, ferromagnetism \cite{Pixley2019,Sharpe2019,Chen2020,Repellin2020}, and other impressive effects \cite{Li2010,Jiang2019,Andrei2021,Slagle2020} just by the appropriate choice of the rotation angle between two layers. The obvious success attracted attention to the Moiré lattices and even led to the rise of the separate field of twistronics.

In this paper, we consider Moiré patterns in the context of nanophotonic metasurfaces. All kinds of photonic crystals and metasurfaces were studied intensively for the last several decades, and their optical properties are known in much detail. In recent years there was a growth of attention to stacks of metasurfaces \cite{fradkin2020thickness,becerril2021optical,murai2021photoluminescence,chen2019empowered,gippius2010resonant,chen2019all,berkhout2020simple,berkhout2019perfect,berkhout2020strong,gerasimov2019engineering} due to the new opportunities that they provide for the control and manipulation of light. Simultaneously, much fewer studies considered the interaction of crystals that do not have a mutual lattice either because of the periods discrepancy or even misalignment of crystallographic axes. The rise of twistronics attracted the deserved attention to this field and gave a significant boost. As a result, currently, there is a number of studies available that demonstrate various effects in photonic Moiré metasurfaces.
Twisted stacks have been used to demonstrate topological transitions of the guided modes \cite{hu2020moire,hu2020topological,zheng2020phonon,duan2020twisted,chen2020configurable,kotov2019hyperbolic,zhou2020polariton}, light localization in Moire supercells \cite{wang2020localization,torrent2020dipolar}, tunable metasurfaces \cite{aftenieva2020tunable}, and chirality enhancement \cite{wu2018high, zhao2012twisted, aftenieva2020tunable}. Also, it has been shown that in the near-field radiative heat transfer between parallel planar metasurfaces \cite{messina2017radiative, kan2019near, dai2015enhanced, fernandez2017enhancing, dai2016near, dai2016radiative}, the heat flux can be controlled by twisting one of the corrugated plates relative to the other \cite{luo2020near, biehs2011modulation,zhou2020polariton,peng2020twist}. 
 Some of the studies utilize laser interference lithography (LIL) to obtain stacks of large-area lattices \cite{aftenieva2020tunable} and single-layer, gradient Moire-metasurfaces \cite{ushkov2020subwavelength,wu2018moire}. Nevertheless, the general understanding of the optical Moiré physics and potential capabilities are still to be understood in the nearest future.

The available calculation approaches for predicting properties of such structures do not provide the full picture as well.
In recent studies, it has been proposed either to apply effective medium approximation \cite{hu2020moire} or to build a transfer/scattering matrix approach for the main (0th) diffraction channel \cite{kotov2019hyperbolic,menzel2016efficient,askarpour2014wave,romain2018study,sperrhake2020equivalence,sperrhake2019analyzing,romain2017spectrally,romain2016extended}. Such approaches undoubtedly boost and simplify the calculations but have strongly limited application area. Effective medium approximation typically describes very dense and thin metasurfaces, whereas the main diffraction order approximation requires the large distance between the sublattices to prevent their interaction via the evanescent waves forming the near field. In this way, none of the approaches allows considering the hybridization of typical quasiguided modes of the twisted photonic structures -- a backbone for plenty of physical systems. 
Universal computational methods such as finite element method (FEM), finite difference time domain one (FDTD), or Fourier modal method (FMM)~\cite{tikhodeev2002}, also known as Rigorous coupled-wave analysis method (RCWA)~\cite{moharam1995}, are not efficient in the application for such structures as well. From one side, the unit cell of the Moiré superlattice is larger (and often much larger) than the unit cells of the original gratings, which significantly complicates the real-space-based calculations (FEM, FDTD). From the other side, a large unit cell corresponds to the small Bragg vector in the reciprocal space and, in turn, a large number of Fourier harmonics have to be accounted for to reach the convergence of reciprocal-space-based FMM.
Remarkably, the most bright twistronic effects are observed for identical materials rotated on very small angles of \cite{Pixley2019,Cao2018,Jiang2019,Sharpe2019,Balents2020}, which corresponds to extremely large periods of the resulting superlattice. This configuration, together with non-twisted sublattices of very close periods, are the most hard-to-compute structures.
In this way, there is a great need for a fast and universal computational method specialized for considering stacks of twisted metasurfaces.

In this paper, we present the adaptation of the Fourier modal method in the form of scattering matrices \cite{tikhodeev2002} for the consideration of twisted stacks of 1D-periodic metasurfaces. Scattering matrices of each sublattice are calculated efficiently, accounting for their 1D-periodic origin, which makes the procedure lean and efficient. To calculate the scattering matrix of the whole stack, we additionally propose to ignore high-$\mathbf{k}_\parallel$ evanescent waves that decay such fast in the gap-layer between sublattices that do not affect the coupling between them at all. Both procedures strongly boost the calculations without noticeable loss of accuracy and pave the way to study any optical properties of the twisted stacks in a reasonable time.
To demonstrate the proposed approach performance, we compare its computation speed with the standard FMM and show the acceleration from 1 up to 3 and more orders of magnitude. Finally, we demonstrate the possible applications on examples of both plasmonic and dielectric twisted stacks. The most difficult for computations plasmonic stack allows controlling the shape of the plasmon dispersion just by a relative rotation of the sublayers. We also show that the chiral diamond photonic crystal slab with inclusions of graphite is able to generate circularly polarized thermal emission (as well as to route the radiation from the circularly polarized emitter) for appropriate twisting angles. The accurate study of these and many other effects become possible with the help of the new approach.

\section{Calculations}

\begin{figure}[h!]
    \centering
    \includegraphics[width=\linewidth]{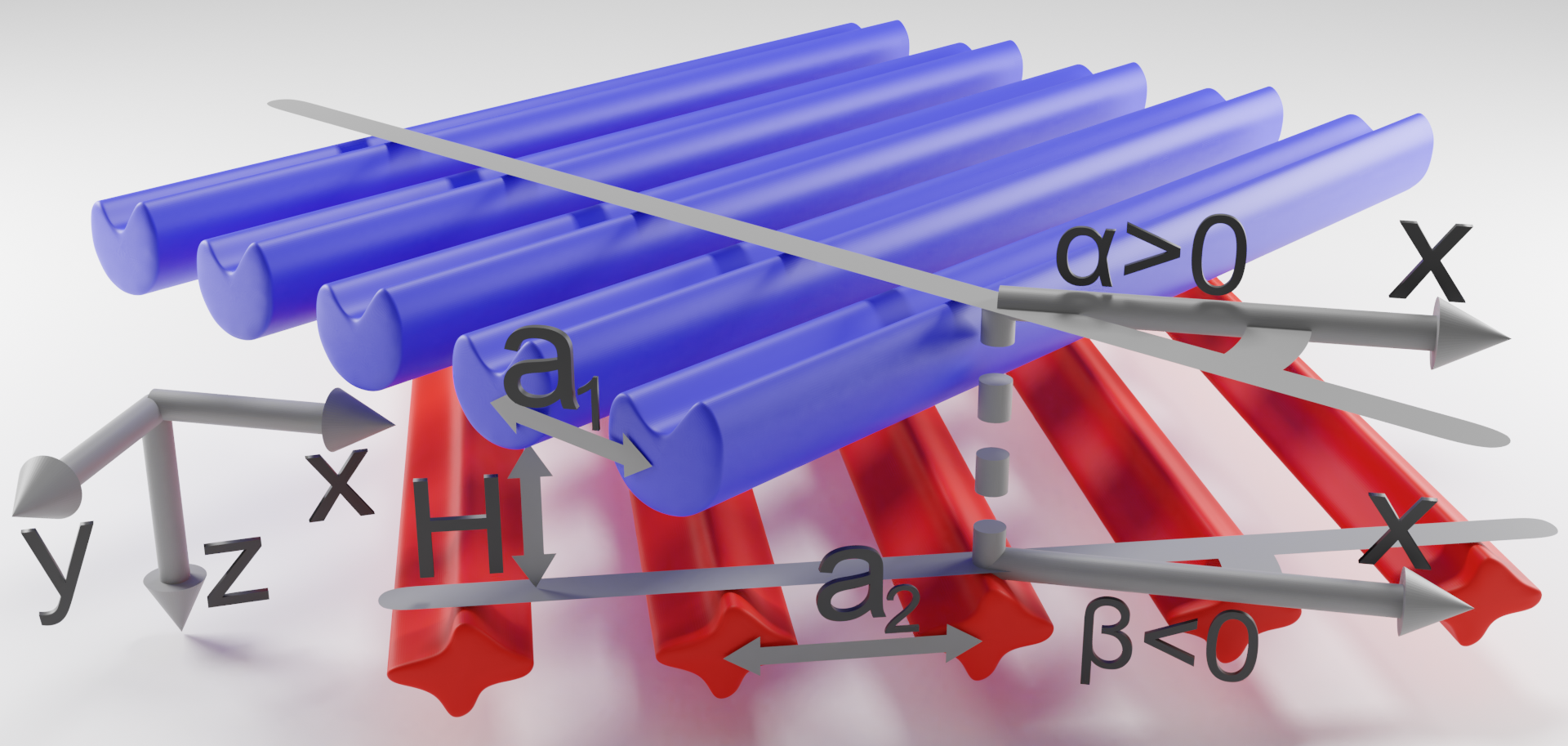}
    \caption{Schematic of the twisted stack consisting of two arbitrary 1D-periodic lattices twisted around a common axis.}
    \label{fig:1}
\end{figure}

\begin{figure*}
    \centering
    \includegraphics[width=0.9\textwidth]{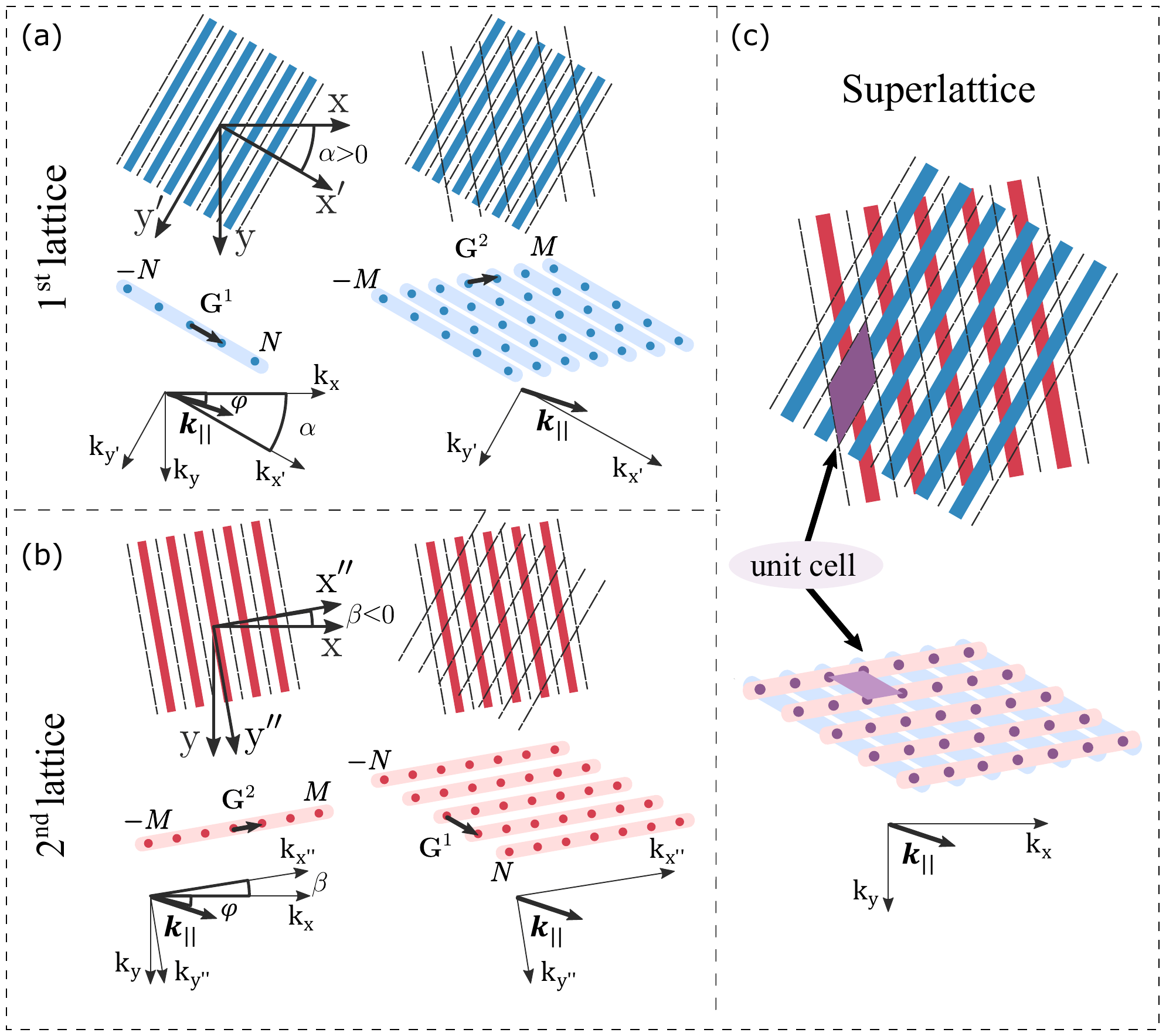}
    \caption{Sketches of the upper (a), lower (b) sublattices and their stack (c) in real and reciprocal spaces. The first column (a,b) schematically shows the structure of the rotated  1D-periodic sublattices and their reciprocal lattices with respect to coordinate axes and wavevector of the incident light $\mathbf{k}_\parallel$. The second column (a,b) formally considers the same sublattices as structures with mutual 2D-periodicity. Each sublattice binds harmonics either of common row or column.
    The last column (c) shows the superimposing of 1D-sublattices resulting in a formation of the true 2D-periodic structure, which substantially connects all the harmonics.
    }
    \label{fig:2}
\end{figure*}

Here, we are going to consider twisted stacks of 1D-periodic gratings that form 2D-periodic Moiré pattern as shown in Fig.~\ref{fig:1}. The upper and lower sublattices might be different, have different periods and arbitrary rotation angles as depicted in Fig.~\ref{fig:1}-\ref{fig:2}. 

Each separate 1D lattice has a corresponding 1D reciprocal lattice (Fig.~\ref{fig:2}~a-b), which is formed by the Bragg vectors $\mathbf{G}^1$ and $\mathbf{G}^2$ that might be imparted by the lattices to the in-plane wavevector of the incident wave $\mathbf{k}_\parallel$. The optical properties of the upper and lower sublattices can be described in terms of the scattering matrices that connect the amplitudes of incoming and outgoing waves.
Nevertheless, the relative rotation and different periods of the lattices result in misalignment and mismatch of their reciprocal lattices (see Fig.~\ref{fig:2}~a-b). This means that each diffraction harmonic of the light passed through one of the lattices generates a new family of harmonics after the diffraction on the second one. Therefore, the structure consisting of upper and lower sublattices is essentially two-dimensional, and its scattering matrix should connect all the corresponding harmonics (see Fig.~\ref{fig:2}~c).

The straightforward approach to calculate spectra of such stacks is just to consider each 1D-periodic sublayer as it is a 2D-periodic metasurface (see the second column in Fig.~\ref{fig:2}~a-b).
For the desired convergence, 2D structures typically require to take into account much more harmonics than 1D ones. Given that the calculation time scales cubically with the number of harmonics~\cite{tikhodeev2002}, this makes 2D calculations relatively slow and resource-consuming. 
The straightforward approach becomes especially inefficient for plasmonic lattices, for which even 1D sublattices require a huge number of harmonics to resolve the high-gradient fields. Brute-force calculation of such structures is challenging even in a single dimension and almost unachievable in 2D.
Moreover, for Moiré lattices, this issue might be even more critical than for typical 2D crystals. As it was already mentioned, the Moiré pattern composed of the sublayers with close vectors of reciprocal lattices $|\mathbf{G}^1-\mathbf{G}^2|\ll k$ (where $k$ is the typical wavevector of light in ambient media) has a large number of opened diffraction channels. Thus, in this case, an achievement of the full convergence for some structures could be especially challenging, which will be shortly discussed on a practical example below.

Nevertheless, we do not consider 2D photonic crystals of the general form, but focus our attention on a rather narrow class of the structures. Utilization of the distinctive features of this class makes it possible to simplify and boost corresponding calculations.

\subsection{Scattering matrix of 1D lattice}

We start with the consideration of the first (upper) 1D sublattice. It is important to note that the derivations will be based on the assumption that the lattice is sandwiched between the two homogeneous layers (see Fig.~\ref{fig:3}). Such environment makes it possible to consider the eigenmodes of the boundary layers as plane linearly-polarized waves, which simplifies the derivations. Although the proposed configuration does not describe the arbitrary structure, we can implement it without losing generality. Indeed, the virtual zero-thickness homogeneous layer can always be inserted, for example, between two back-to-back gratings.

Let us separate out the layer bounded by horizontal dashed lines (see Fig.~\ref{fig:2}(a), Fig.~\ref{fig:3}) that includes upper sublattice in 
specially aligned with it $x'-y'$ coordinate system. Looking ahead, the similar layer for the lower sublattice should be chosen standing back-to-back with the first one, i.e. the gap interlayer is virtually divided in an arbitrary ratio between the layers. 
Incident light determines the in-plane components of the main harmonic wavevector $\mathbf{k}_{\parallel}=k_{x}\hat{x}+k_{y}\hat{y}=k_{x'}\hat{x'}+k_{y'}\hat{y'}$, where $\hat{x}$, $\hat{y}$, $\hat{x'}$ and $\hat{y'}$ are unit vectors defining the direction of the corresponding axes. Interaction of light with lattice leads to an emergence of diffraction harmonics that have corresponding wavevectors that might be arranged in a hypervectors:

\begin{multline}
    \begin{bmatrix}\vec{K}^1_{x'} &\vec{K}^1_{y'}\end{bmatrix}  =
    \begin{bmatrix}K^{1,-N}_{x'} &K^{1,-N}_{y'}\\
    K^{1,-N+1}_{x'} &K^{1,-N+1}_{y'}\\
    \vdots\\
    K^{1,0}_{x'} &K^{1,0}_{y'}\\
    \vdots\\
    K^{1,N}_{x'} &K^{1,N}_{y'}\end{bmatrix}  =\\= \begin{bmatrix}1\\1\\\vdots\\1\\\vdots\\1\end{bmatrix}\begin{bmatrix}k_{x'}\\ k_{y'}\end{bmatrix}^\mathrm{T}+\begin{bmatrix}-N\\-N+1\\\vdots\\0\\\vdots\\N\end{bmatrix}\begin{bmatrix}G^1_{x'}\\G^1_{y'}\end{bmatrix}^\mathrm{T},
\end{multline}
where $G^1_{x'}$, $G^1_{y'}$ are projections of the Bragg vector on denoted axes and the number of harmonics that are taken into account is $N_g=2N+1$. The finite number of these harmonics makes it possible to build the scattering matrix that connects the amplitudes of incoming $ [\vec{d}'_-,\vec{u}'_+]^T$ and outgoing $    [\vec{d}'_+,\vec{u}'_-]^T$ waves on the boundaries of the considered layer (horizontal dashed lines in Fig.~\ref{fig:3}). Since the consideration is held in the primed basis, we consider the matrix $\mathbb{S}^{'}_{1}(\omega,\mathbf{k}_\parallel)=\mathbb{S}^{'}_{1}(\omega,[k_{x'},k_{y'}]^\mathrm{T})$ as a function of $k_{x'}$ and $k_{y'}$ projections:

\begin{figure}
    \centering
    \includegraphics[width=0.95\linewidth]{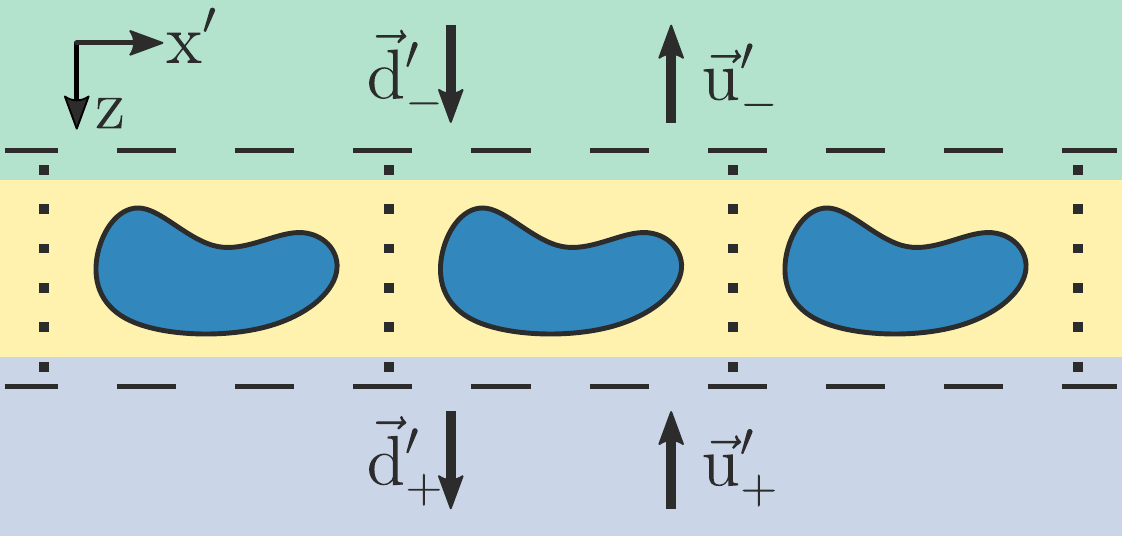}
    \caption{Scheme of the 1D lattice sandwiched between homogeneous adjacent layers. Long dashed lines separate the periodic part of the structure from the outer dielectric environment and lie in homogeneous layers, short dashed lines denote the boundaries of the unit cells. Arrows illustrate the incoming and outgoing harmonics that are connected via the scattering matrix.}
    \label{fig:3}
\end{figure}

\begin{equation}
    \begin{bmatrix}
    \vec{d}'_+\\\vec{u}'_-
    \end{bmatrix}=
    \mathbb{S}^{'}_{1}(\omega,[k_{x'},k_{y'}]^\mathrm{T})
    \begin{bmatrix}
    \vec{d}'_-\\\vec{u}'_+
    \end{bmatrix},
\end{equation}
where $\vec{u'_\pm}$ and $\vec{d'_\pm}$ are hypervectors of amplitudes correspond to the upward and downward propagating waves, $\pm$ signs indicate the section that has a larger and smaller $z$-coordinate correspondingly.
Each of the hypervectors $\vec{u}'_{\pm},\vec{d}'_{\pm}$ describes $2N_g$ modes of different polarizations, whereas scattering matrix $\mathbb{S}'_1$ has $4N_g\times4N_g$ size respectively.
If the boundaries of the considered layer are located inside a homogeneous environment (the case which we consider here), then according to our definition, amplitudes correspond to $x'-$ and $y'-$polarized plane waves:

\begin{equation}
    \vec{u}'_{\pm}=\begin{bmatrix}\vec{u}_{\pm,x'}\\\vec{u}_{\pm,y'}\end{bmatrix},\quad     \vec{d}'_{\pm}=\begin{bmatrix}\vec{d}_{\pm,x'}\\\vec{d}_{\pm,y'}\end{bmatrix},
\end{equation}
where the amplitudes are sorted in accordance with hypervectors $[\vec{K}^1_{x'},\vec{K}^1_{y'}]$ in such a way that the field of up- and downgoing waves can be found as follows:

\begin{equation}
    \begin{bmatrix}
    E_{\pm,x'}(\mathbf{r}_{\parallel}) \\ E_{\pm,y'}(\mathbf{r}_{\parallel})
    \end{bmatrix}=\sum_{n=-N}^{N} \left(\begin{bmatrix}u^{n}_{\pm,x'}\\u^{n}_{\pm,y'}
    \end{bmatrix}+\begin{bmatrix}d^{n}_{\pm,x'}\\d^{n}_{\pm,y'}
    \end{bmatrix}\right)e^{i\mathbf{K}^{1,n}_\parallel\mathbf{r}_\parallel},
\end{equation}
where index $n$ indicates $n$-th component of the hypervectors, $\mathbf{K}_\parallel^{1,n} = K_x^{1,n}\hat{x}+K_y^{1,n}\hat{y}$ and $\mathbf{r}_\parallel = x\hat{x}+y\hat{y}$.

\subsection{Scattering matrix of the rotated sublattice}

Scattering matrix $\mathbb{S}_1^{'}$ fully describes the optical properties of the considered layer and can be found by standard FMM realisations. This matrix is a function of frequency $\omega$ and in-plane component of the wavevector $\mathbf{k}_\parallel$, but also depends on the structure's parameters. The scattering matrix's transformation with the rotation of the lattice can be easily found and does not require some additional computations. To demonstrate this, we consider the first lattice rotated by an angle $\alpha$ and find the connection of scattering matrix in $x-y$ coordinate system $\mathbb{S}_1(\omega,\mathbf{k}_\parallel)=\mathbb{S}_1\left(\omega,[k_{x},k_{y}]^\mathrm{T}\right)$ with matrix $\mathbb{S}'_1(\omega,\mathbf{k}_\parallel)=\mathbb{S}_1\left(\omega,[k_{x'},k_{y'}]^\mathrm{T}\right)$ 
defined in $x'-y'$ basis (see Fig.~\ref{fig:2}~a). The latter one has already been considered:

\begin{equation}
    \begin{bmatrix}
    d_{+,x'}\\d_{+,y'}\\u_{-,x'}\\u_{-,y'}
    \end{bmatrix}=
    \mathbb{S}^{'}_1\left(\omega,[k_{x'},k_{y'}]^\mathrm{T}\right)
     \begin{bmatrix}
    d_{-,x'}\\d_{-,y'}\\u_{+,x'}\\u_{+,y'}
    \end{bmatrix}.
\end{equation}
Hypervectors that are connected by the scattering matrix might be easily expressed in the original $x-y$ coordinate system:
\begin{multline}
    \begin{bmatrix}
    d_{+,x}\\d_{+,y}\\u_{-,x}\\u_{-,y}
    \end{bmatrix}=\mathbb{R}_\alpha       \begin{bmatrix}
    d_{+,x'}\\d_{+,y'}\\u_{-,x'}\\u_{-,y'}
    \end{bmatrix},\quad
 \begin{bmatrix}
    d_{-,x}\\d_{-,y}\\u_{+,x}\\u_{+,y}
    \end{bmatrix}=\mathbb{R}_\alpha    \begin{bmatrix}
    d_{-,x'}\\d_{-,y'}\\u_{+,x'}\\u_{+,y'}
    \end{bmatrix},\\\mathbb{R}_\alpha=
    \begin{bmatrix}
    \cos(\alpha)\hat{I}&-\sin(\alpha)\hat{I}&\hat{0}&\hat{0}\\
    \sin(\alpha)\hat{I}&\cos(\alpha)\hat{I}&\hat{0}&\hat{0}\\
    \hat{0}&\hat{0}&\cos(\alpha)\hat{I}&-\sin(\alpha)\hat{I}\\
    \hat{0}&\hat{0}&\sin(\alpha)\hat{I}&\cos(\alpha)\hat{I}\\
    \end{bmatrix}, 
\end{multline}
where $\mathbb{R}_\alpha$ is the rotation matrix for the hypervectors, $\hat{I}$ is $N_g\times N_g$ identity matrix and $\hat{0}$ is $N_g\times N_g$ zero matrix. 
Due to the fact that both scattering matrices describe the same mapping in different bases, we can easily obtain the final expression for the scattering matrix in the original coordinates:
\begin{equation}
    \mathbb{S}_1(\omega,[k_{x},k_{y}]^\mathrm{T}) = \mathbb{R}_{\alpha}    \mathbb{S}^{'}_1(\omega,\hat{R}_{-\alpha}[k_{x},k_{y}]^\mathrm{T})\mathbb{R}_{-\alpha},
\end{equation}
where
\begin{equation}
    \begin{bmatrix}
    k_{x'}\\k_{y'}
    \end{bmatrix}=\hat{R}_{-\alpha} 
    \begin{bmatrix}
    k_{x}\\k_{y}
    \end{bmatrix},\quad \hat{R}_\alpha=
    \begin{bmatrix}
    \cos(\alpha)&-\sin(\alpha)\\
    \sin(\alpha)&\cos(\alpha)
    \end{bmatrix}.
\end{equation}

We should remember that the in-plane components of the wavevectors of the diffraction harmonics, $\mathbf{K}^{1,n}_\parallel=K^{1,n}_{x}\hat{x}+K^{1,n}_{y}\hat{y}=K^{1,n}_{x'}\hat{x'}+K^{1,n}_{y'}\hat{y'}$, are transformed accordingly:

\begin{multline}
    \begin{bmatrix}
        K^{1,n}_{x}(\mathbf{k}_\parallel)\\
        K^{1,n}_{y}(\mathbf{k}_\parallel)
    \end{bmatrix} = 
    \begin{bmatrix}
    k_{x}\\k_{y}    
    \end{bmatrix} +n 
    \begin{bmatrix}
    G^1_{x}\\G^1_{y}    
    \end{bmatrix}= \\\hat{R}_{\alpha}\begin{bmatrix}
        K^{1,n}_{x'}(\mathbf{k}_\parallel)\\
        K^{1,n}_{y'}(\mathbf{k}_\parallel)
    \end{bmatrix}= \hat{R}_{\alpha}    \begin{bmatrix}
    k_{x'}\\k_{y'}    
    \end{bmatrix} +n \hat{R}_{\alpha}
    \begin{bmatrix}
    G^1\\0    
    \end{bmatrix},
\end{multline}
where $G^1$ is the absolute value of the Bragg vector of the first lattice (${G^1}=G^1_{x'}=\sqrt{{G_x^1}^2+{G_y^1}^2}$).

The same procedure holds for the second lattice, which is rotated on an angle $\beta$ and is aligned with the double-primed axes:

\begin{equation}
    \mathbb{S}_2(\omega,[k_{x},k_{y}]^\mathrm{T}) = \mathbb{R}_{\beta}    \mathbb{S}^{''}_2(\omega,\hat{R}_{-\beta}[k_{x},k_{y}]^\mathrm{T})\mathbb{R}_{-\beta}.
\end{equation}
The corresponding basis is the following:

\begin{equation}
    \begin{bmatrix}
        K^{2,m}_{x}(\mathbf{k}_\parallel)\\
        K^{2,m}_{y}(\mathbf{k}_\parallel)
    \end{bmatrix} = 
    \begin{bmatrix}
    k_{x}\\k_{y}    
    \end{bmatrix} +m 
    \begin{bmatrix}
    G^2_{x}\\G^2_{y}    
    \end{bmatrix} = \hat{R}_{\beta}    \begin{bmatrix}
    k_{x''}\\k_{y''}    
    \end{bmatrix} +m \hat{R}_{\beta}
    \begin{bmatrix}
    G^2\\0    
    \end{bmatrix}.
\end{equation}
The number of harmonics for this case is denoted $M_g=2M+1$.

\subsection{Scattering matrix of the stack in mutual basis }

We have obtained the scattering matrices of the rotated 1D metasurfaces, but our final goal is to construct the scattering matrices derived in a mutual basis, which requires us to attribute them formally as 2D crystals. The mutual, reciprocal lattice for both sublattices might be found as a direct sum of their own 1D reciprocal lattices. In practical calculations, we take into account only a finite number of harmonics, and the easiest approach is also to take the direct sum of finite sublattices as shown in Fig.~\ref{fig:2}~(a-b), second column. Such an approach is not the only one, but it is highly convenient for practical implementation.

The wavevectors of the 2D superlattice harmonics might be arranged in a single hypervector in different orders. The straightforward approach is to place wavevectors of the 2D lattice "row" after "row" or "column" after "column." In other words, harmonics of the mutual basis can be constructed as a concatenation of sets of harmonics of 1D lattices for different wavevectors of the main harmonic.

For the consideration of the first (upper) lattice, it is convenient to arrange the wavevectors in the order, which we indicate by the letter $\mathrm{A}$ (see the second column of Fig.~\ref{fig:2}~(a-b)):

\begin{multline}
    \begin{bmatrix}{}^\mathrm{A}\vec{K}^{\mathrm{2D}}_x &{}^\mathrm{A}\vec{K}^{\mathrm{2D}}_y\end{bmatrix}=
    \\ \begin{bmatrix}\vec{K}^1_{x}(\mathbf{k_\parallel}-M\mathbf{G}^2) &\vec{K}^1_{y}(\mathbf{k_\parallel}-M\mathbf{G}^2)\\
    \vec{K}^1_{x}(\mathbf{k_\parallel}+(-M+1)\mathbf{G}^2) &\vec{K}^1_{y}(\mathbf{k_\parallel}+(-M+1)\mathbf{G}^2)\\
    \vdots&\vdots\\
    \vec{K}^1_{x}(\mathbf{k_\parallel}+M\mathbf{G}^2) &\vec{K}^1_{y}(\mathbf{k_\parallel}+M\mathbf{G}^2)\end{bmatrix} =\\
    \begin{bmatrix}    \vec{K}^{1}_{x}\left(\mathbf{K}^{2,-M}_{\parallel}[\mathbf{k}_\parallel]\right)&\vec{K}^{1}_{y}\left(\mathbf{K}^{2,-M}_{\parallel}[\mathbf{k}_\parallel]\right)
    \\
    \vec{K}^{1}_{x}\left(\mathbf{K}^{2,-M+1}_{\parallel}[\mathbf{k}_\parallel]\right)&\vec{K}^{1}_{y}\left(\mathbf{K}^{2,-M+1}_{\parallel}[\mathbf{k}_\parallel]\right)
    \\
    \vdots&\vdots
    \\
    \vec{K}^{1}_{x}\left(\mathbf{K}^{2,M}_{\parallel}[\mathbf{k}_\parallel]\right)&\vec{K}^{1}_{y}\left(\mathbf{K}^{2,M}_{\parallel}[\mathbf{k}_\parallel]\right)
    \\\end{bmatrix},
\end{multline}
where the expression $\vec{K}^1_{x/y}(\mathbf{q})$ indicates that the corresponding hypervector is calculated for the wavevector of the main harmonic equal to $\mathbf{q}$.

With such a choice of the basis, each block of the scattering matrix of the first lattice (considered as 2D one) splits into sub-blocks that describe the internal coupling of each "row"'s harmonics. Since the lattice is fundamentally one-dimensional, all the inter-"row" connections are equal to zero, which makes sparse the scattering matrix ${}^\mathrm{A}\mathcal{S}_1$ of the first lattice formally considered as 2D crystal. To demonstrate this, we first represent the scattering matrices in the form, where each block is responsible for either reflection or transmission of $x$- or $y$-polarized incident light to the $x$- or $y-$ polarized channels:

\begin{multline}
    {}^\mathrm{A}\mathcal{S}_1(\omega,\mathbf{k}_\parallel) = \left(\begin{array}{c|c|c|c}
        {}^\mathrm{A}\mathcal{S}_1^{11} &    {}^\mathrm{A}\mathcal{S}_1^{12}    &    {}^\mathrm{A}\mathcal{S}_1^{13} &    {}^\mathrm{A}\mathcal{S}_1^{14}\\
    \hline
            {}^\mathrm{A}\mathcal{S}_1^{21} &    {}^\mathrm{A}\mathcal{S}_1^{22}    &    {}^\mathrm{A}\mathcal{S}_1^{23} &    {}^\mathrm{A}\mathcal{S}_1^{24}\\
    \hline
            {}^\mathrm{A}\mathcal{S}_1^{31} &    {}^\mathrm{A}\mathcal{S}_1^{32}    &    {}^\mathrm{A}\mathcal{S}_1^{33} &    {}^\mathrm{A}\mathcal{S}_1^{34}\\
    \hline
            {}^\mathrm{A}\mathcal{S}_1^{41} &    {}^\mathrm{A}\mathcal{S}_1^{42}    &    {}^\mathrm{A}\mathcal{S}_1^{43} &    {}^\mathrm{A}\mathcal{S}_1^{44}\\
    \end{array}\right),\\
    \mathbb{S}_1(\omega,\mathbf{k}_\parallel) = \left(\begin{array}{c|c|c|c}
        \mathbb{S}_1^{11} &    \mathbb{S}_1^{12}    &    \mathbb{S}_1^{13} &    \mathbb{S}_1^{14}\\
    \hline
            \mathbb{S}_1^{21} &    \mathbb{S}_1^{22}    &    \mathbb{S}_1^{23} &    \mathbb{S}_1^{24}\\
    \hline
            \mathbb{S}_1^{31} &    \mathbb{S}_1^{32}    &    \mathbb{S}_1^{33} &    \mathbb{S}_1^{34}\\
    \hline
            \mathbb{S}_1^{41} &    \mathbb{S}_1^{42}    &    \mathbb{S}_1^{43} &    \mathbb{S}_1^{44}\\
    \end{array}\right),\label{eqn:16blocks}
\end{multline}
where blocks ${}^\mathrm{A}\mathcal{S}_1^{ij}$ has $N_gM_g\times N_gM_g$ size, whereas blocks $\mathbb{S}_1^{ij}$  are $N_g\times N_g$ size correspondingly. In turn, the blocks are obviously connected as follows:
\begin{multline}
    {}^\mathrm{A}\mathcal{S}_1^{ij}(\omega,\mathbf{k}_\parallel) =\\ \begin{pmatrix}\mathbb{S}_1^{ij}(\mathbf{K}^{2,-M}_{\parallel}[\mathbf{k}_\parallel])&\hat{0}&\dots&\hat{0}\\
    \hat{0}&\mathbb{S}_1^{ij}(\mathbf{K}^{2,-M+1}_{\parallel}[\mathbf{k}_\parallel])&\dots&\hat{0}\\
    \vdots&\vdots&\ddots&\vdots\\
    \hat{0}&\hat{0}&\dots&\mathbb{S}_1^{ij}(\mathbf{K}^{2,M}_{\parallel}[\mathbf{k}_\parallel])
    \end{pmatrix}.
\end{multline}

Next, we need to build the scattering matrix of the second lattice ${}^\mathrm{A}\mathcal{S}_2$ in the same basis $\mathrm{A}$. However, it is hard to do it once and therefore, we first consider the scattering matrix of the second lattice in its natural basis denoted B (see the "chains" in the second column of Fig.~\ref{fig:2}~b):
\begin{multline}
    {}^\mathrm{B}\mathcal{S}_2^{ij}(\omega,\mathbf{k}_\parallel) =\\ \begin{pmatrix}\mathbb{S}_2^{ij}(\mathbf{K}^{1,-N}_{\parallel}[\mathbf{k}_\parallel])&\hat{0}&\dots&\hat{0}\\
    \hat{0}&\mathbb{S}_2^{ij}(\mathbf{K}^{1,-N+1}_{\parallel}[\mathbf{k}_\parallel])&\dots&\hat{0}\\
    \vdots&\vdots&\ddots&\vdots\\
    \hat{0}&\hat{0}&\dots&\mathbb{S}_2^{ij}(\mathbf{K}^{1,N}_{\parallel}[\mathbf{k}_\parallel])
    \end{pmatrix}.
\end{multline}
where the harmonics are sorted as follows:
\begin{multline}
    \begin{bmatrix}{}^\mathrm{B}\vec{K}^{\mathrm{2D}}_x &{}^\mathrm{B}\vec{K}^{\mathrm{2D}}_y\end{bmatrix}=\\
    \vec{K}^2_{x}(\mathbf{k_\parallel}+N\mathbf{G}^1)
    \begin{bmatrix}    \vec{K}^{2}_{x}\left(\mathbf{K}^{1,-N}_{\parallel}[\mathbf{k}_\parallel]\right)&\vec{K}^{2}_{y}\left(\mathbf{K}^{1,-N}_{\parallel}[\mathbf{k}_\parallel]\right)
    \\
    \vec{K}^{2}_{x}\left(\mathbf{K}^{1,-N+1}_{\parallel}[\mathbf{k}_\parallel]\right)&\vec{K}^{2}_{y}\left(\mathbf{K}^{1,-N+1}_{\parallel}[\mathbf{k}_\parallel]\right)
    \\
    \vdots&\vdots
    \\
    \vec{K}^{2}_{x}\left(\mathbf{K}^{1,N}_{\parallel}[\mathbf{k}_\parallel]\right)&\vec{K}^{2}_{y}\left(\mathbf{K}^{1,N}_{\parallel}[\mathbf{k}_\parallel]\right)
    \\\end{bmatrix}.
\end{multline}

The block-wise computation of the matrices must be faster than a straightforward approach. Indeed, the calculation time of each scattering matrix scales cubically with the number of harmonics. Therefore, the naive consideration of each sublayer as 2D crystal takes $\tau_{\mathrm{straightforward}}\propto N_g^3M_g^3$, whereas our block-wise-calculation approach for the first and second lattices scales as $\tau_{1}\propto N_g^3M_g$ and $\tau_{2}\propto N_gM_g^3$, which is much faster in practice, especially for the large number of harmonics.

The knowledge of ${}^\mathrm{A}\mathcal{S}_1(\omega,\mathbf{k}_\parallel)$ and ${}^\mathrm{B}\mathcal{S}_2(\omega,\mathbf{k}_\parallel)$ matrices is still not enough to obtain the stack scattering matrix. To do that, we need to consider matrices of both sublayers in a mutual basis. In our case, bases A and B differ only by the order of their elements and, therefore, one of them can be reduced to another by simple rearrangement. The transition matrix that connects them can be easily derived.

Indeed, when we consider the basis A, then the (n,m) harmonic (see Fig.~\ref{fig:2}~a) has the index number $p_\mathrm{A} = (m+M)N_g+n+N+1$. The same harmonic in the basis B (see Fig.~\ref{fig:2}~b) has the index number $p_\mathrm{B}=(n+N)M_g+m+M+1$. In this terms the matrices in $\mathrm{A}$ and $\mathrm{B}$ representations are connected in a very simple way:

\begin{equation}
    {}^{\mathrm{A}}\mathcal{S}^{ij}_{2,p_\mathrm{A}q_\mathrm{A}} = {}^{\mathrm{B}}\mathcal{S}^{ij}_{2,p_\mathrm{B}q_\mathrm{B}},
\end{equation}
where indexes $ij$ indicate one of sixteen blocks (see~Eqn.~\ref{eqn:16blocks}) and the indices $p_\mathrm{A/B}q_\mathrm{A/B}$ run inside each of the blocks.

In practice, there are several possible ways to implement the described permutation. The most obvious is just to iterate over all the indices in a loop to find the scattering matrix components. Nevertheless, much more effective approach is to build once the mappings $p_\mathrm{A}=T_\mathrm{AB}(p_\mathrm{B})$ and $p_\mathrm{B}=T_\mathrm{BA}(p_\mathrm{A})$ between the representations in advance and 
apply it on demand. $T_\mathrm{AB}$ and $T_\mathrm{BA}$ are primitive functions tabulated according to the described rule.
In this way, we need just to apply the permutation operation, which is available in most packages for numerical computations.

\begin{equation}
    {}^{\mathrm{A}}\mathcal{S}^{ij}_{2,p_\mathrm{A}q_\mathrm{A}} = {}^{\mathrm{B}}\mathcal{S}^{ij}_{2,T_\mathrm{BA}(p_\mathrm{A})T_\mathrm{BA}(q_\mathrm{A})}.
\end{equation}
The resulting matrix ${}^\mathrm{A}\mathcal{S}_2$ is still sparse but is no longer block-diagonal, which results in the "mixing" of blocks related to the upper and lower sublattices and arise of specific Moiré-induced physical effects.

Once we get scattering matrices of both sublayers in a mutual basis $\mathrm{A}$ (or any other one) the total scattering matrix of the whole stack might be found  ${}^\mathrm{A}\mathcal{S}={}^\mathrm{A}\mathcal{S}^1\otimes{}^\mathrm{A}\mathcal{S}^2$via the well-known formulas~\cite{Weiss2009} (from here on we omit index $A$ implying that all the matrices are derived in the same basis):

\begin{align}
    \mathcal{S}^{\mathrm{dd}}&=    \mathcal{S}_2^{\mathrm{dd}}\left(\hat{I}-\mathcal{S}_1^{\mathrm{du}}\mathcal{S}_2^{\mathrm{ud}}\right)^{-1}\mathcal{S}_1^{\mathrm{dd}},\label{eqn:comb1}
\\
    \mathcal{S}^{\mathrm{du}}&=   \mathcal{S}_2^{\mathrm{du}}+ \mathcal{S}_2^{\mathrm{dd}}\left(\hat{I}-\mathcal{S}_1^{\mathrm{du}}\mathcal{S}_2^{\mathrm{ud}}\right)^{-1}\mathcal{S}_1^{\mathrm{du}}\mathcal{S}_2^{\mathrm{uu}},\label{eqn:comb2}
\\
    \mathcal{S}^{\mathrm{ud}}&=   \mathcal{S}_1^{\mathrm{ud}}+ \mathcal{S}_1^{\mathrm{uu}}\left(\hat{I}-\mathcal{S}_2^{\mathrm{ud}}\mathcal{S}_1^{\mathrm{du}}\right)^{-1}\mathcal{S}_2^{\mathrm{ud}}\mathcal{S}_1^{\mathrm{dd}},\label{eqn:comb3}
\\
    \mathcal{S}^{\mathrm{uu}}&=    \mathcal{S}_1^{\mathrm{uu}}\left(\hat{I}-\mathcal{S}_2^{\mathrm{ud}}\mathcal{S}_1^{\mathrm{du}}\right)^{-1}\mathcal{S}_2^{\mathrm{uu}},\label{eqn:comb4}
\end{align}
where the matrices are considered in terms of large $2N_gM_g\times2N_gM_g$ blocks:

\begin{align}
    \mathcal{S}(\omega,\mathbf{k}_\parallel)& = \left(\begin{array}{c|c}
        \mathcal{S}^{dd} &    \mathcal{S}^{du}\\
        \hline\mathcal{S}^{ud} &    \mathcal{S}^{uu}
    \end{array}\right),\\
        \mathcal{S}_1(\omega,\mathbf{k}_\parallel)& = \left(\begin{array}{c|c}
        \mathcal{S}_1^{dd} &    \mathcal{S}_1^{du}\\
        \hline\mathcal{S}_1^{ud} &    \mathcal{S}_1^{uu}
    \end{array}\right),\\
    \mathcal{S}_2(\omega,\mathbf{k}_\parallel)& = \left(\begin{array}{c|c}
        \mathcal{S}_2^{dd} &    \mathcal{S}_2^{du}\\
        \hline\mathcal{S}_2^{ud} &    \mathcal{S}_2^{uu}
    \end{array}\right).    
\end{align}

Each of the formulas \ref{eqn:comb1}-\ref{eqn:comb4} contains the inversion of the sparse matrices for which, according to our knowledge, there are no effective specialized algorithms. In this way, since the time of the matrix inversion scales cubically with its size, the total computation time for the stack scattering matrix has the same cubic asymptotic $\tau\propto N^3_gM^3_g$ as the naive approach. Nevertheless, even the lean calculation of each sublattice matrix, as we will show below, accelerates the whole procedure multiple times.

\section{Filtration of harmonics}

In most practical cases, calculation of the Moiré stack scattering matrix might be additionally boosted as well. If there is a thin but finite homogeneous gap-layer of thickness $H$ (see Fig.~\ref{fig:1}) in-between twisted metasurfaces, then additional physical-based simplification can be implemented. Indeed, high-$k_\parallel$ evanescent waves (requiring a large number of harmonics in the calculation to be accounted for) might be needed to accurately describe the optical properties of separate upper or lower lattice. Nevertheless, these harmonics decay in the gap in such a fast way that they do not make almost any contribution in their interaction. Therefore, it is easy to formulate the criterion of taking harmonic into account for the calculation of matrices combination:

\begin{equation}
    e^{-\mathrm{Im}\left(\sqrt{\varepsilon_{\mathrm{gl}}\omega^2/c^2-(\mathbf{k}_\parallel+n\mathbf{G}^1+m\mathbf{G}^2)^2}H\right)}>\Theta,\label{eqn:condition}
\end{equation}
where $\varepsilon_{\mathrm{gl}}$ is the permittivity of the gap-layer and $\Theta$ is the cut-off level, which determines the tolerance of the fully converged results and should be in the range of $10^{-2}$ -- $10^{-10}$ for most of the structures. The fulfillment of this condition results in throwing away a significant number of corresponding rows and columns from the matrices $\mathbb{S}_1$ and $\mathbb{S}_2$ and subsequent drastic increase of  Eqns.~\ref{eqn:comb1}-\ref{eqn:comb4} calculation speed. Moreover, the number of remaining harmonics saturates rapidly, making the computation time of matrices combination almost constant as well. In this case, the asymptotic time-scaling will be determined by the sublattices $\tau\propto \max(N_g^3M_g,N_gM_g^3)$. If someone is not satisfied with the $\Theta$-limited precision, it is possible to reduce the cut-off level synchronously with the increase of the number of harmonics. This would worsen the asymptotic but guarantee the ultimate precision with a remaining great gain in speed.

It is important that despite the filtration, we still account for the nearfield of the metasurfaces and the potential coupling of their guided modes by the nearfield interaction. It is important to specially emphasize that although high harmonics are not in demand at the finishing stage, accounting for the interaction between the sublayers, they are vital to obtaining the low-$\mathbf{k}_\parallel$ cores of $\mathbb{S}_1$ and $\mathbb{S}_2$ matrices since sublattices might efficiently "mix" all the harmonics. Therefore, the number of harmonics $N_gM_g$ used for $\mathbb{S}_1$ and $\mathbb{S}_2$ calculation should be in general significantly larger than the number, which fulfil the condition~\ref{eqn:condition}.

In the case of zero or extremely small $H$, the filtering technique is not applicable and does not increase the calculation efficiency. However, this case corresponds to tightly bound lattices that form a true 2D crystal, and it is reasonable that the simplification based on the concept of complex independent structures connected via the limited number of channels is irrelevant.

\subsection{Further development}

Computational approaches presented in the previous sections provide a fast and efficient calculation of twisted Moiré stacks spectra, which will be demonstrated below. Nevertheless, the computations can be further improved, which might be vital in most complicated cases.

The most obvious improvement is associated with the filtration of harmonics. In most practical cases we keep rather limited number of low-$\mathbf{k}_\parallel$ harmonics that  lie in a circle defined by the cut-off level in Eqn.~\ref{eqn:condition}. Without loss of generality, if one consider the upper lattice then it becomes obvious that most of the "chains" of the harmonics (second column of Fig.~\ref{fig:2}~(a)) are out of any relation to the low-$\mathbf{k}_\parallel$ of our interest. Therefore corresponding blocks of the scattering matrices can be excluded, which additionally improves the time calculation asymptotic $\tau\propto \max(N_g^3,M_g^3)$. Such approach requires a little bit more complicated procedure for the choice of the mutual basis and harmonics of individual lattices, which should account for the cut-off level. Nevertheless, this procedure is straightforward and might be easily realised in the spirit of the technique presented in the study. Potential interest for efficient calculations might be attracted by the choice of other alternative 2D harmonics sets that are not direct sums of any other 1D harmonics sets.

One more opportunity to improve the computations for the weakly-coupled metasurfaces lies in a simplified accounting for their interaction. Indeed, it such cases it is most likely possible to develop some sort of perturbation theory. Apparently, it will be associated with the substitution of expressions with inversion of matrices with series expansion (see Eqns.~\ref{eqn:comb1}-\ref{eqn:comb4}). Potentially, the use of only sparse matrix multiplication, which is a fast operation, might significantly boost the calculations. Nevertheless, this approach requires separate detailed consideration and rigorous study of its potential applicability.

Yet another acceleration can be applied for the hard-to-compute Moiré superlattices of large Moiré-periods. As it was already discussed, similar sublattices superimposition results in a formation of very dense reciprocal lattice, and according to the described procedure, one needs to calculate the blocks of the matrices that connect very closely-located chains of harmonics. Therefore, corresponding scattering matrices might be found not directly, but from the interpolation of precalculated matrices on a mesh in reciprocal space. Such approach, not only strongly reduces the number of repeated calculations, but also helps to choose the optimal for the problem set of harmonics (not necessarily direct-sum-generated).

\section{Performance}

\begin{figure}[h!]
    \centering
    \includegraphics[width=0.95\linewidth]{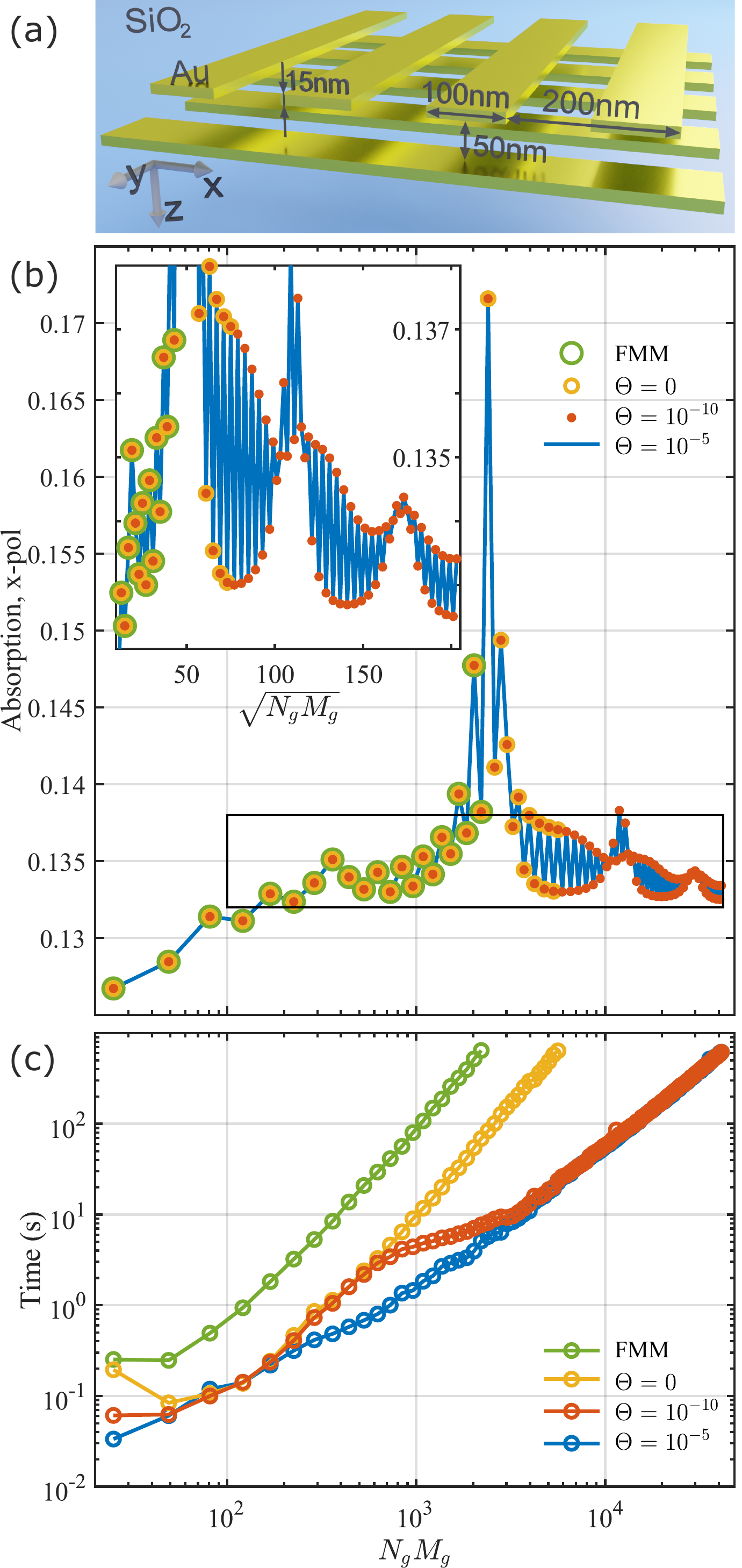}
    \caption{Study of the computations convergence with the number of harmonics used to calculate the absorption of normally-incident $x$-polarized light in the plasmonic stack depicted in panel (a). Panel (b) shows the dependence of the absorption on the total number of harmonics taken into account in a logarithmic scale. The inset employs the linear scale for the square root of harmonics to demonstrate the details of the oscillations near the expected limit. Panel (c) shows the calculation time scaling with the number of harmonics for the different approaches. Green lines and markers correspond to the reference FMM, yellow ones to our approach without filtration, whereas red and blue ones correspond to different filtration cut-off levels.}
    \label{fig:4}
\end{figure}

In the previous sections, we have proposed an effective computational approach and estimated its asymptotic time consumption. Nevertheless, it is important to observe the acceleration in practice. To do that, we have chosen one of the most hard-to-compute structures - the stack of two 1D plasmonic lattices in silica ($\varepsilon_{\mathrm{SiO}_2}=1.46^2$). Fig.~\ref{fig:4}~(a) indicates the dimensions of perpendicular golden lattices that are identical to each other. The optical properties of gold are described by Johnson\&Christy optical constants~\cite{JohnsonChristy1972}.

As an illustration, we consider the normal incidence of x-polarized 1200-nm light on the structure. Fig.~\ref{fig:4}~(b) and (c) show the dependence of the absorption and the calculation time, respectively, on the number of accounted Fourier harmonics correspondingly (we take $N_g=M_g$ for all the computations). Green lines and markers are responsible for the application of the standard Fourier-modal method (enhanced by Li's factorization rules \cite{Li1997new,Li1996}) to a 2D grating (straightforward approach) and act as reference values.
One can see from Fig.~\ref{fig:4}~(c) that in logarithmic axes, single-processor calculation time almost immediately goes to a straight line of almost constant slope $\approx2.5$, which is slightly smaller than the theoretical asymptotic limit of $3$. The last computed point corresponding to $47^2=2209$ Fourier harmonics takes approximately $600$ seconds
. 
Nevertheless, due to the peculiarities of the metal properties, the 
absorption value is still far from convergence even for such a large number of harmonics as $2209$ (see Fig.~\ref{fig:4}).

The application of the techniques proposed in this paper allows us to accelerate the calculations significantly. For the zero cut-off level (yellow lines and markers in Fig.~\ref{fig:4}~(b-c)) there is no filtration of high-$\mathbf{k}_\parallel$ harmonics at all, which results in the same asymptotic time scaling (panel (c)) and the identical absorption (panel (b)), which matches the standard FMM up to machine precision. Nevertheless, the optimal approach for calculating the sublayers scattering matrices makes the computations approximately one order of magnitude faster, which is clearly seen from the gap between the parallel green and yellow lines (panel (c)). It is just because the number of the most asymptotically expensive mathematical operations is reduced approximately tenfold. This gives the possibility to account for a much larger number of harmonics at the same time.

\begin{figure*}[t!]
    \centering
    \includegraphics[width=\linewidth]{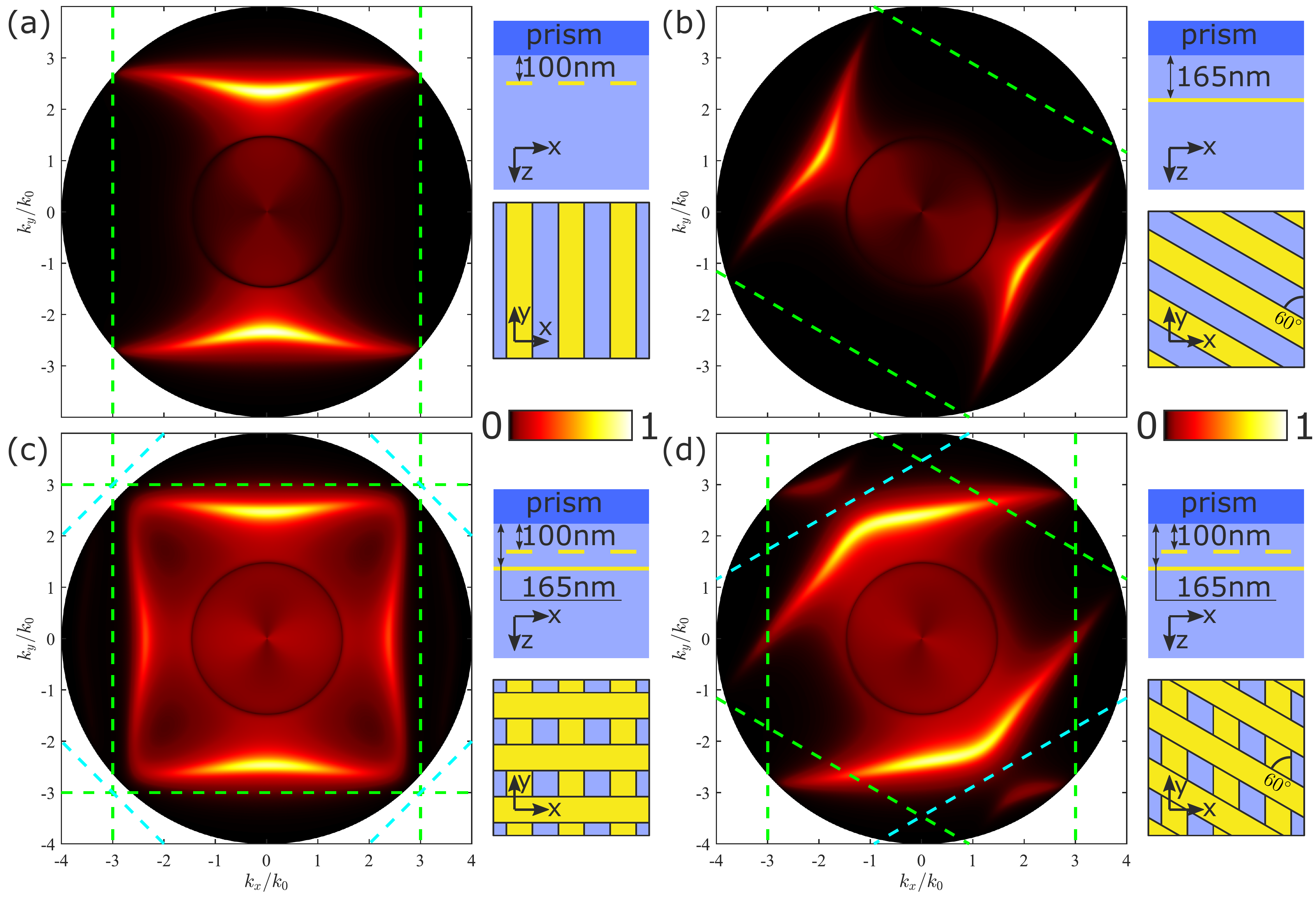}
    \caption{Absorption maps of $p$-polarised $1200$~nm-wavelength light in four sketched $\mathrm{SiO}_2$-embedded plasmonic lattices illuminated through the prism ($x-y$ projections show the bottom-up view, which is in accordance with other figures). Panels (a-b) demonstrate the hyperbolic-like behavior of plasmons in solitary lattices placed at different distances from the prism. Panels (c-d) show the modes hybridization in stacks of (c) perpendicularly- and (d) $60\degree$-angle-crossed identical plasmonic lattices. Green dashed lines indicate the boundaries of the 1D  lattices Brillouin zones, whereas the cyan lines are Moiré-pattern-associated boundaries. }
    \label{fig:5}
\end{figure*}

Nevertheless, even more promising results are demonstrated by the approximate method that filters out high Fourier harmonics on the stage of calculating the combination of sublayers scattering matrices. 
Even such a small cut-off level as $\Theta=10^{-10}$ results in a total change of the computation time dependence (red line in panel (c)). Indeed, while the number of harmonics is relatively small, none of them is ignored (Fig.~\ref{fig:4}~(c), red and yellow lines coincide), but as soon as their number overcomes the threshold of approximately 670 harmonics, time to compute scattering matrices combination stops growing. This results in a bend of the graph line and flattening of its asymptotic. As it was shown before, theoretical estimation gives the quadratic scaling of the computation time $\tau\propto(N_gM_g)^2$ (for $N_g=M_g$), but in practice, we see the power of approximately $\approx1.7$.
In this way, the harmonics filtration results in an additional strong acceleration of the computations, and, moreover, this acceleration becomes the greater, the more harmonics are taken into account.
It is also important that although the filtration-based approach is approximate, the computed absorption values match perfectly with the previous results, and no significant deviation is seen (see Fig.~\ref{fig:4}~(b)). Therefore, the level of the filtration-specified precision does not prevent obtaining almost the full convergence potentially.

A relatively high cut-off level of $\Theta=10^{-5}$ (blue lines and markers) shifts the threshold of line-bending to a much smaller number of harmonics so that it is no longer distinguishable (panel (c)). As it is seen from the graphs, this gives us a significant acceleration for the middle number of harmonics (300-3000), which can be most practical
in calculations for many structures. At the same time, the coincidence with the previous, accurate calculations is perfect, which proves the remaining high level of precision (panel (b)).

The optimal choice of the number of harmonics and the cut-off level significantly depends on the materials and geometry of the structure, physical effects observed in the considered frequency range, and obviously on the desired precision.

\section{Examples}

The high speed and precision of computations makes it much easier to compute spectra of Moiré metasurfaces. In particular,
here we explore the potential applications of the computational approach on the examples of the already discussed plasmonic lattices (twisted on different angles) and membrane diamond photonic crystal slab.

\subsection{Plasmonic crystal}

\begin{figure*}[t!]
    \centering
    \includegraphics[width=\linewidth]{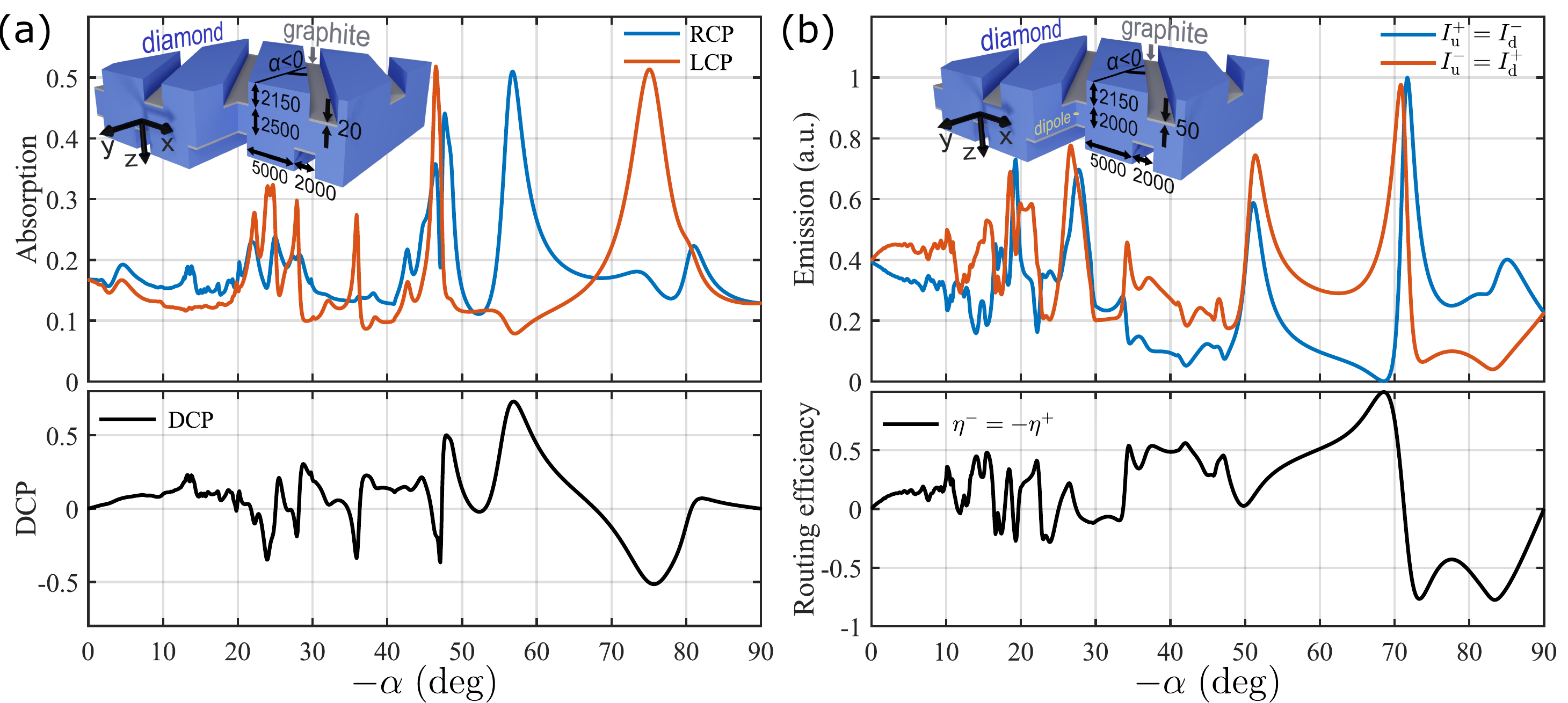}
    \caption{Absorption (a) and emission (b) of $10\mu$m-light from the twisted diamond photonic crystal slab with graphite inclusions (insets). Panel (a) shows the absorption of the right- and left-handed circularly-polarized light as well as the degree of circular polarisation in absorption as a function of the rotation angle. Panel (b) demonstrates the angle-dependence of the upwards and downwards emission in the normal directions for the circularly polarised $\sigma^\pm$ dipoles. The bottom figure shows the routing efficiencies $\eta^-=-\eta^+$, which almost reach the ultimate unitary value. 
    }
    \label{fig:6}
\end{figure*}

Plasmonic crystals and as well as inclusions of plasmonic nanoparticles in photonic crystal slabs are well known for their ability to localize light at the nanoscale and form hybrid high-$Q$ collective resonances~\cite{gippius2010resonant,fradkin2019fourier,fradkin2020nanoparticle,fradkin2020thickness,zundel2021lattice,cuartero2020super,baur2018,utyushev2021collective,guo2017,bin2021ultra,reshef2019multiresonant,Yermakov2018,samusev2017polarization,kolkowski2019lattice,vaskin2019light,kostyukov2021multipolar,gerasimov2021plasmonic,guo2019lasing,nevcada2020multiple}. Without a doubt, they can become the basis for a variety of Moiré superlattices as well.

The natural way to study the optical properties of the twisted plasmonic gratings is to consider isofrequency dispersion curves.
In order to do that, we employ auxiliary high-index optical prism ($\varepsilon_{\mathrm{prism}}=16$) at a small distance from the structure (see schemes in Fig.~\ref{fig:5}) and use it for excitation of the guided modes. The absorption maps for $p-$polarized $1200$~nm light, which excite surface plasmons in the plasmonic structures, are shown in Fig.~\ref{fig:5}. Panels (a) and (b) demonstrate that solitary lattices (constituents of the up-following stacks) support plasmons propagating along the golden sheets. Indeed, the longitudinally polarized surface waves that propagate along the structure almost do not feel the vertical walls of nanostrips and behave similarly to plasmons in an an infinite metal slab. At the same time, the plasmonic modes polarized perpendicularly to the strips faces strong depolarization due to subwavelength dimensions of the latter one. For this reason such modes are naturally associated with localized resonances that are observed for smaller wavelengths. The fact that at the considered wavelength plasmon waves can propagate in one direction and cannot in the other one leads to 
concavity of its dispersion.
An effective anisotropic medium can not accurately describe these plasmons due to the proximity of the first Brillouin zone boundaries (green dashed lines) to dispersion curves. Nevertheless, dispersion curves demonstrate typical hyperbolic-like behavior~\cite{yermakov2015hybrid}, which is widely spread in various nanophotonic applications~\cite{hu2020moire}. It is worth noting that the modes of the structure located 100~nm from the prism (panel (a)) are obviously excited stronger than the ones of the 165~nm-disposed structure (panel (b)) due to decay of evanescent waves in $\mathrm{SiO}_2$ layer.

Hybridization of upper- and lower-lattice plasmons in a stack is able to change their dispersion significantly.
Interestingly, even interaction of perpendicularly crossed lattices leads to a formation of the closed square-like dispersion curve (see Fig.~\ref{fig:5}~(c)). This effect is very similar to a topological transition in twisted graphene metasurfaces~\cite{hu2020moire} and probably also can be implemented for so-called field canalization \cite{belov2005canalization,hu2020moire}. Nevertheless, similar to the solitary lattices (panels a-b) $x$-propagating plasmons are less pronounced than $y$-propagating ones, which is explained just by an asymmetric prism-assisted excitation.

Relative rotation of the lattices is a powerful tool to shape the dispersion in a desired way. The $60\degree$ angle between nanoribbons makes the first Brillouin zone (Wigner–Seitz cell) a regular hexagon (see Fig.~\ref{fig:5}~(d)), whereas the structure remains $C_2$-symmetric. Two pairs of its boundaries shown by green dashed lines originate from the reciprocal lattices of separate 1D sublattices, whereas the cyan pair corresponds to a specific Moiré-inspired effect associated with both sublattices.
The interaction of lattices is most pronounced, where the resonances of individual lattices intersect (superimposition of panels (a) and (b)). Indeed, we observe strong anticrossing of the surface plasmons near the "cyan" boundary of the Brillouin zone. Nevertheless, in this example, hybridization is mostly due to the interaction through the main harmonic. The lattice effect associated with a vicinity of the Brillouin zone boundary and corresponding interaction with Bragg-vector-coupled harmonics is very weak and does not significantly affect the behavior of the structure. 
The shape of the newly-formed dispersion curve is obviously determined by the original plasmons but strongly differs form them at the same time. Most importantly, dispersion becomes convex, which totally changes the behavior of the plasmon wavepackage propagation.

\subsection{Dielectric crystal}

As an example of dielectric Moiré structure, we explore optical properties of diamond photonic crystal slabs (see insets in Fig.~\ref{fig:6}) in the mid-infrared frequency range ($\lambda=10 \mu$m). These structures are diamond membranes in which identical periodic series of grooves are cut from both sides. The bottom of the grooves is covered by a thin, several-dozen-nanometer layer of graphite, which is naturally formed during the direct laser writing~\cite{komlenok2019optical}. The grooves are twisted on an angle $\alpha$ with respect to each other (the lower one is aligned with $y$ axis, $\beta=0$, whereas the upper is rotated by the angle, $\alpha<0$); the exact dimensions of the structures are indicated in Fig.~\ref{fig:6}. Diamond and graphite permittivities at 10$\mu$m wavelength are $\varepsilon_{d} = 5.6581+0.0004i$ and $\varepsilon_{g} = 9.99 + 35.55i$ correspondingly~\cite{komlenok2019optical}.

The considered structures do not have any mirror symmetry in the general case, which makes them chiral and perspective for manipulating circularly polarized light. In this way, we illuminate the structure by normally incident light of complementary circular polarizations and observe the absorption, which occurs predominantly (but not completely) in graphite. As one can see from Fig.~\ref{fig:6}~(a), absorption strongly depends on the angle $\alpha$ - there are lots of resonances that are excited for different angles.
We see that the degree of circular polarisation, $\mathrm{DCP}=(A_{\mathrm{RCP}}-A_{\mathrm{LCP}})/(A_{\mathrm{RCP}}+A_{\mathrm{LCP}})$, for the absorption reaches the high level of approximately 73\% for approximately 57$\degree$ angle of rotation (see the lower graph in Fig.~\ref{fig:6}~(a)). Even higher levels can be achieved with a thinner layer of graphite, but the finite physically meaningful thickness "blurs" the pure effect and limits its maximal value. According to the Lorentz reciprocity principle, the shown effect means that the thermal emission of a $10\mu$m-wavelength, which can be observed from the Moiré structure, will be strongly circularly polarized in the normal direction. In this way, we can consider this structure as a passive source of circularly polarized thermal emission, which might be potentially implemented in the radiative heat transfer problems~\cite{luo2020near,zhou2020polariton,peng2020twist}.

As for the angular dependence, it is clear that relatively large angles such as 50-90$\degree$ correspond to typical 2D lattices, which have comparable to each other periods in two direction. Periods of the structure, as well as its optical properties, slowly change with an angle variation in this range, and therefore the resonances are wide in terms of the angular width. Nevertheless, as we discussed hereinbefore, one of the structure periods tends to infinity with an angle going to zero. This results in a large number of sharp resonances and Rayleigh anomalies that interchange each other in the range of $15-50\degree$. At the same time, the efficiency of the high harmonics excitation rapidly falls with angle decrease, and as a result, the amplitude of the dense modulation in Fig.~\ref{fig:6}~(a) becomes negligible for $0-15\degree$. It is important to emphasize once again that small angles of rotation potentially require taking into account a huge number of harmonics at least to cover open diffraction channels. Nevertheless, in practice, we observe that up to a high level of precision, the convergence is achieved even when not all the opened channels are accounted for, which indirectly proves the weak contribution of the ignored harmonics to the properties of the whole structure. This effect paves the way to the potential studies of even large Moiré-period superlattices.

We consider the chiral structure, which has not only $z$-axis possessing a 2-fold rotation, but also two other axes of the same kind lying in the $x-y$ plane. Overall, the structure has $D_2$ point group symmetry, making it perspective for the routing of the radiation from the circularly polarized dipole emitter~\cite{Dyakov2020}. Indeed, let us consider a point dipole that is located in the structure according to the inset of Fig.~\ref{fig:6}~(b) - in the middle of each of the graphite strips (in terms of $x-y$ plane projection) and in the center of the slab (in terms of the $z$ coordinate). We denote the dipoles of complementary circular polarisations $\mathbf{P}^{\pm}=[1,\pm i,0]^\mathrm{T}$ as $\sigma^\pm$ correspondingly. In this way, due to the symmetry of the structure, the intensity of light emitted upwards at a normal angle by $\sigma^+$ dipole $I^+_u$ is equal to the intensity of the $\sigma^-$ dipole downwards emission $I^-_d=I^+_u$. The same relation is valid for the complementary quantities $I^-_u=I^+_d$. This peculiarity simplifies the rout of $\sigma^\pm$ dipoles emission to opposite directions. In case we manage to nullify the downwards emission of $\sigma^-$ dipole $I^-_d=0$ that would automatically make the upwards emission of $\sigma^+$ dipole zero as well $I^+_u=0$, which means that each dipole will radiate in its own direction. Such routing effect in $D_4$-symmetric structure was already studied in detail in our previous paper~\cite{Dyakov2020}.

It is convenient to evaluate the emission directivity numerically via the efficiencies $\eta^\pm=\frac{I^\pm_u-I^\pm_d}{I^\pm_u+I^\pm_d}$, which are connected with each other as well $\eta^+=-\eta^-$. The accurate choice of the geometric parameters allowed us to achieve almost perfect routing (more than 99.5\% efficiency) for approximately $68\degree$ angle of rotation (Fig.~\ref{fig:6}~(b)). It is clear from the upper graph in Fig.~\ref{fig:6}~(b) that the most strong effect corresponds to the nullification of the emission in one of the directions due to the Fano lineshape of the resonance (blue line). In principle, the structure might be optimized to demonstrate this effect for narrow small-angle resonances that are observed in emission as well.

\section*{Conclusion}

In this paper, we have proposed a modification of the Fourier modal method for consideration of twisted 1D lattices that form Moiré patterns. We have demonstrated the possibility of speeding up the computations from 1 to 3 and more orders depending on the number of harmonics without a noticeable decrease in the accuracy. Such an approach paves the way for rigorous study of modes hybridization in twisted photonic crystals slabs and the potential design of optical devices based on them. As an example, we have demonstrated the tuning of plasmonic modes dispersion and in a stack of plasmonic lattices. Also, the utilization of the diamond photonic crystal slab for circularly polarized thermal emission in the mid-infrared range and the emission routing from the circularly-polarized dipole source have been shown.

\section*{Acknowledgements}
This work was supported by the Russian Foundation for
Basic Research (Grant No. 20-02-00745).

N.S.K and I.M.F. contributed equally to this work. N.S.K and I.M.F. developed the program code and prepared the manuscript. S.A.D. and N.A.G. supervised the study. All authors contributed to the discussions and commented, reviewed, and edited on the paper.


%

\end{document}